\documentclass{svjour2}
\usepackage{eurosym}
\usepackage{graphicx}

\smartqed
\begin{document}

\title{Quantum decoherence: a logical perspective}
\author{Sebastian Fortin \and Leonardo Vanni }
\institute{ Sebastian Fortin  \at CONICET, Department of Physics, FCEN (UBA), Argentina. \email{sfortin@gmx.net}        \and
           Leonardo Vanni \at Department of Physics, FCEN (UBA), Argentina. \email{Lvanni@df.uba.ar}
}

\maketitle

\begin{abstract}
The so-called classical limit of quantum mechanics is generally studied in
terms of the decoherence of the state operator that characterizes a system.
This is not the only possible approach to decoherence. In previous works we
have presented the possibility of studying the classical limit in terms of
the decoherence of relevant observables of the system. On the basis of this
approach, in this paper we introduce the classical limit from a logical
perspective, by studying the way in which the logical structure of quantum
properties corresponding to relevant observables acquires Boolean
characteristics.

\end{abstract}


\institute{ Sebastian Fortin  \at CONICET, Department of Physics, FCEN (UBA), Argentina. \email{sfortin@gmx.net}        \and
           Leonardo Vanni \at Department of Physics, FCEN (UBA), Argentina. \email{Lvanni@df.uba.ar}
}




\section{Introduction}

There are different perspectives to address the problem of the classical
limit of quantum mechanics. The orthodox treatment introduces the phenomenon
of decoherence as the key to solve this problem \cite{Bub1997}. The
mainstream approach to decoherence is the so-called \textquotedblleft
environment induced decoherence\textquotedblright , developed by Zurek and
his collaborators (see, e.g., \cite{Zurek1981} \cite{Zurek2003} \cite{ZurekPaz2002}).
In the context of this approach, the goal is to know
whether the state becomes diagonal or not \cite{Schlosshauer2007}. If the
state becomes diagonal, then it acquires the structure of a mixed state of
classical mechanics; this feature leads to the usual interpretation of the
decohered state from a classical viewpoint.

In our group, we have developed a general theoretical framework for
decoherence based on the study of the evolution of the expectation values of
certain relevant observables of the system \cite{GTFD}. According to this
framework, decoherence is a phenomenon relative to the relevant observables
selected in each particular case \cite{EPSA2009}.

This new approach and the orthodox
treatment of decoherence are equivalent from a mathematical point of view
(see \cite{Formal}). Nevertheless, there are good reasons to think that the
treatment of decoherence by means of the behavior of the observables of the
system instead of that of its states may have conceptual advantages. For
example, it allow us to study the decoherence in closed systems like the
universe \cite{SID7}, it does not have the problem of distinguishing between
system and environment \cite{EPSA2009} and this allows us to study open
systems without using the reduced state \cite{fortinlombardi2014}%

The purpose of this work is to argue that the main advantage of the study of
decoherence in terms of the Heisenberg representation is that this approach
allows us to analyze the logical aspects of the classical limit. On the one
hand, we know that the lattices of classical properties are distributive or
Boolean \cite{BirkhoffNeumann1936}: when operators are associated with those
properties, they commute with each other. On the other hand, it is well
known that the lattices of quantum properties are non-distributive, a formal
feature manifested by the existence of non-commuting observables  \cite{Bub1997} \cite{CohenLibro}.
 In spite of this difference, there are certain
quantum systems which, under certain particular conditions, evolve in a
special way: although initially the commutator between two operators is not
zero, due to the evolution it tends to become zero \cite{KieferPolarski2009}%
. Therefore, in these systems it should be possible to show that, initially,
they can be represented by a non-Boolean lattice, but after a definite time
a Boolean lattice emerges: this process, that could be described from the
perspective of the expectation values of the system's observables, deserves
to be considered as a sort of decoherence that leads to the classical limit.
In other words, from this perspective the classical limit can be addressed
by studying the dynamical evolution of non-Boolean lattices toward Boolean
lattices. In the present work we will study this transition from the
viewpoint of the general theoretical framework for decoherence.

\section{Classical and Quantum Logic}

The logical structure of a theory can be studied from the set of properties
the theory is able to describe, more specifically, by analyzing under what
circumstances an isomorphism between the set of properties and the sentences
of language that predicates those properties can be established. If the
isomorphism can be consistently established, then the sentences of the
language correspond to the properties, and the logical operations on
language sentences correspond to certain algebraic operations on the
corresponding properties.

The mathematical structure of the properties is not exactly the same as the
logical structure of the theory, but by studying the first it is possible to
determine the second. The structure of the sentences (propositions) of
language can be read and analyzed from the mathematical structure of
properties.

When we speak about the properties of the theory, at least in the case of a
physical theory, we have to consider the `\emph{value properties}'
associated with the physical quantities that the theory describes. So, if
an observable  $\hat{O}$   of  a physical system, with values $o_{i}$, acquires the value $o_2$ when
the system is in certain conditions (in some state, say $\varphi $), then a
value property is represented by the pair definite by  $p_{2}=$ `$(\hat{O}:o_{2})$', and the
corresponding sentence could be expressed as $L_{2}=$ `\emph{when the system
is in state} $\varphi ,$ \emph{the magnitude} $\hat{O}$ \emph{has} \emph{the
value} $o_{2}$ '.

The simplest mathematical structure that can be studied is established with  a
\emph{partial order relation} between  properties. A partial order, $\preceq $,
is an order relation satisfying reflexivity, transitivity and antisymmetry
\cite{CohenLibro}. The order relation between the properties is closely related
with the logical entailment between sentences in language. However, not all ordering
relations at the level of the properties can be linked to a well-defined
entailment. Entailment faces the problem of defining a truth function on  of properties (essentially, a function that assigns values $0$ and $1$ to the properties)
 that can give rise to a logical structure of sentences. This task can be non trivial in the quantum case \cite{MittelstaedtLibroLogica}.

However, even without a well-defined truth function, it is possible to
establish a \emph{probability function} on  of properties. A
probability function $\mathcal{P}$ is a function evaluated on the set of the
properties $P$, which assigns a value between zero and one: $\mathcal{P}%
:P\rightarrow \lbrack 0,1]$. In this case, the link between sentences and
properties is established in probabilistic terms. Then, a property
represented by $p_{2}=$ `$(O:o_{2})$' may correspond to a sentence $L_{2}=$ `%
\emph{when the system is in the state} $\varphi ,$ \emph{the magnitude $O$}
\emph{has the value} $o_{2}$ \emph{with probability} $\mathcal{P}(p_{2})=0.2$%

Endowed with a order relation between  properties it is possibile to define the  operations
\emph{meet} $\wedge $, \emph{join} $\vee $, and \emph{complement} $\perp $.
These operations respectively correspond to the usual logical connectives
between the sentences of the language: conjunction, disjunction and negation
\cite{Bub1997} \cite{HughesLibro}.
The meet between two properties is
defined as the \emph{greatest lower bound }$(GLB)$ between them, and the
join as the \emph{least upper bound} $(LUB)$ \cite{CohenLibro}. In turn, the
complement $p^{\perp }$ of a property $p$ satisfies $p\wedge p^{\perp }=0$
and $p\vee p^{\perp }=1$.

When the meet and the join exist  for all pairs of the properties, then this
 defines a \emph{lattice} of properties $R=(%
\emph{P},\preceq )$, where $\emph{P}$ is the set of all properties, and $%
\preceq $ is an order relation  \cite{CohenLibro} \cite{HughesLibro}.

If $\perp $ further satisfies
\begin{eqnarray}
&&{(p^{\perp })}^{\perp }=p  \nonumber \\
&&p\preceq q\Rightarrow q^{\perp }\preceq p^{\perp }  \nonumber
\end{eqnarray}%
then the lattice is said \textit{orthocomplemented}.

The lattice   of properties, with the  operations meet, join,
and complementation   representing logical connectives between the corresponding sentences of the language, determines  an algebraic structure of properties.
This structure characterizes the logical aspects of the theory and allows us to study them.

In the classical case, the set of properties corresponding to the sentences
of the language is determined by all the possible subsets of the phase space
of the system, and the partial order relation is given by the inclusion
between sets. This leads to a representation of the logical conjunction,
disjunction, and negation in the classical discourse by means of the typical
operations of intersection, union and complement between sets \cite{Bub1997}%
. The resulting structure determines a Boolean algebra \cite%
{OmmesLibroInterpretacion}; it is usually said that classical lattices are
Boolean lattices \cite{Boole1854}.

The quantum case is very different. The set of quantum properties is
determined by the closed subspaces of the Hilbert space of the system under study
\cite{BirkhoffNeumann1936}. This fact introduces crucial differences in the
definition of the operations representing the logical connectives, and has
peculiar consequences in the structure of the quantum discourse.

The partial order relation between properties is given by the inclusion of
subspaces of Hilbert space. The meet operation is still the intersection,
but now between subspaces. The differences are introduced in the join and
complementation operations. The join between two properties of a quantum
lattice is defined by the closure of the subspace spanned by the linear combinations of the
elements of the subspaces representing such properties. That is to say, it
is the space spanned by the subspaces of each property \cite{HughesLibro}.
Finally, the complementation of a property is given by the orthogonal
complement of the subspace representing that property.

As we have already pointed out, although it is not always possible to
establish a well-defined truth function on the lattice of properties, it is
nevertheless possible to define a probability function on it, although with
certain limitations. Not all probability functions satisfy the axioms of
Kolmogorov. The differences among them depend on the Boolean structure (or
not) of the lattice \cite{MittelstaedtLibroLogica}. It can be proved that
only on Boolean lattices it is possible to introduce a probability function
well defined in the Kolmogorovian sense \cite{CohenLibro} \cite{HughesLibro} \cite{Holik}  \cite{Redei} \cite{Gudder}.
 In more general lattices, as quantum lattices, the probability function is
well defined (in the Kolmogorovian sense)  only when it is applied on Boolean sublattices.

A simple form of encoding the logical differences between quantum and
classical lattices consists in analyzing the validity of the so-called
`distributive equalities' \cite{CohenLibro}. The distributive equalities
express the distributivity of the operation meet with respect to the
operation join, and vice versa. However, these equalities are not always
valid. In general, only distributive inequalities hold. Given the properties
$a$, $b$ and $c$, the following inequalities are always valid

\begin{eqnarray}
a\wedge (b\vee c) &\succeq &(a\wedge b)\vee (a\wedge b)  \nonumber \\
a\vee (b\wedge c) &\preceq &(a\vee b)\wedge (a\vee b)  \nonumber
\end{eqnarray}%
Only in a Boolean lattice the equalities hold.

Another important aspect associated with distributive inequalities is that
they capture the notion of \emph{compatibility} as understood in quantum
mechanics \cite{CohenLibro}. It can be proved that, if the properties $a$
and $b$ are such that 
\begin{eqnarray}
a &=&(a\wedge b)\vee (a\wedge b^{\perp })  \nonumber \\
b &=&(b\wedge a)\vee (b\wedge a^{\perp })  \nonumber
\end{eqnarray}%
then the projectors associated with the subspaces representing these
properties commute. Otherwise, the projectors do not commute and their value
properties are incompatible. Of course, if those projectors are involved in
the spectral decomposition of the observables $A$ and $B$, then $A$ and $B$
are also incompatible observables.

We arrive thus to an important conclusion. Only when all the properties to
be described are associated with compatible observables, there is a Boolean
structure corresponding to a classical description, and in this case the
distributive equalities hold and the probabilities are well defined in
Kolmogorovian sense. Otherwise, there are incompatible observables, the
lattice structure is not Boolean, and the probabilities, in general, are not
well defined in the Kolmogorovian sense.

\section{Incompatibility of observables in time}

As it is well known, quantum mechanics admits at least two representations.
The Schr\"{o}dinger representation studies the evolution of the state $\hat{%
\rho}(t)$, and the Heisenberg representation studies the evolution of the
observables $\hat{O}(t)$ \cite{Sakurai}. The traditional approach to
decoherence emphasizes the evolution of the state in the Schr\"{o}dinger
representation: it studies the diagonalization of the state in the preferred
basis \cite{Zurek1982} \cite{ZurekPaz2002}. Such diagonalization removes
interference, which is one of the phenomena specific of quantum mechanics.
However, this approach does not make explicit the disappearance of another
peculiar feature of quantum mechanics, that is, contextuality.   Contextuality is
linked to the non-commutativity of observables, because two non-commuting
observables belong to different contexts. The Heisenberg uncertainty
principle is other manifestation of non-commutativity, and expresses the
fact that it is not possible to simultaneously measure the value of two
non-commuting observables.  This principle establishes a fundamental difference with classical mechanics,
where all the observables commute with each other. Therefore, any attempt to
construct a classical limit should include a mechanism capable of explaining
the transition from non-commutativity to commutativity.

In the Schr\"{o}dinger picture, if a pair of observables do not commute in
the initial time,
\[
\left[ \hat{O}_{1},\hat{O}_{2}\right] \neq 0
\]%
then they do not commute ever, since observables do not evolve. For this
reason, the natural picture to study the transition from non-commutativity
to commutativity is the Heisenberg picture. Some authors, like Kiefer and
Polarski, described decoherence in the Heisenberg representation
\cite{KieferPolarski2009}, \cite{KieferPolarski1998} . In the present paper, our
purpose is to continue this line of work by studying the time evolution of
the logical properties of quantum systems. Our aim is to find a process in
which two observables do not commute at the initial time, but they do
commute later:

\[
\left[ \hat{O}_{1}(0),\hat{O}_{2}(0)\right] \neq0\longrightarrow\left[ \hat{O%
}_{1}(t),\hat{O}_{2}(t)\right] \cong0
\]

For this purpose, we will use the approach to decoherence called
'`Self-Induced Decoherence' (SID), developed in our group \cite{GTFD}  \cite{Formal} \cite{SID7} \cite{SID1} \cite%
{SID2} \cite{SID3} \cite{SID4} \cite{SID5} \cite{SID6}  \cite%
{SID8} \cite{SID9} \cite{SID10} \cite{SID11} \cite{SID12} \cite{SID13}  \cite{SIDDiscreto}  \cite{Brasilero}. This approach will
allow us to easily show the process of interest. Nevertheless, the same
result can be obtained by the orthodox `Environment Induced Decoherence'
approach.

\subsection{Self-Induced decoherence in the Heisenberg picture}

Although at present EID is still considered the "orthodoxy" in the subject
\cite{Bub1997}, other approaches have been proposed to face its problems, in
particular, the closed-systems problem. One of them is SID, according to
which a closed quantum system with continuous spectrum may decohere by
destructive interference and reach a final state where the classical limit
can be rigorously obtained \cite{SID7} \cite{SID1} \cite%
{SID2} \cite{SID3} \cite{SID4} \cite{SID5} \cite{SID6}  \cite%
{SID8} \cite{SID9} \cite{SID10} \cite{SID11} \cite{SID12} \cite{SID13}  \cite{SIDDiscreto}  \cite{Brasilero}.

Self-Induced decoherence (SID) is a formalism that finds its roots in an
algebraic formalism which was initiated in the Brussels school, lead by
Prigogine \cite{Antoniou}. In this paper we will use the notation according
to which the observables are thought as vectors, and we write them as $\hat{O%
}=\left\vert O\right) $. SID considers a closed quantum system governed by a
Hamiltonian with continuous spectrum. Then we can write a generic observable
as $\hat{O}=\int \int O\left( \omega \omega ^{\prime }\right) \left\vert
\omega \omega ^{\prime }\right) d\omega d\omega ^{\prime }$ where $O\left(
\omega \omega ^{\prime }\right) $ is a generic distribution, and $\left\vert
\omega \omega ^{\prime }\right) $ are generalized eigenvectors of space
observable, that is, $\left\{ \left\vert \omega \omega ^{\prime }\right)
\right\} $ is the basis of space. This notation is necessary for technical
reasons we will not discuss in this article but can be found in \cite{Antoniou}.

According to the work of our group, the different approaches to decoherence
can be described from a General Theoretical Framework for Decoherence
consisting of 3 steps \cite{GTFD} \cite{Formal} \cite{SIDDiscreto} \cite%
{Brasilero}. The most important step is to choose a subset of the
observables of interest.

EID adopts the open system prespective, that is,
the relevant observables have the form $\hat{O}_{R}=\hat{O}_{S}\otimes \hat{I%
}_{E}$, where $\hat{I}_{E}$ is the identity in the space of the environment
and $\hat{O}_{S}$ is any observable of the proper system. On the other hand
SID selects the van Hove observables as relevant observables, it is a good
choice because the restriction on the observables does not diminish the
generality of this approach, because the observables not belonging to the
van Hove space are not experimentally accessible. Then, if we compute the
evolution of the mean values of the van Hove observables, we find that the
interference terms disappear. Then it is possible to understand this prosses
as a kind of decoherence \cite{GTFD}. You can find an exhaustive comparison
between SID and EID in the paper \cite{Brasilero}. Here we present a version
of Self-Induced decoherence in the Heisenberg picture.

 In this case, we will select special observables
that are appropriate to our study, i.e., the time evolution of commutators.
In SID, we consider a quantum system with Hamiltonian $H$ with continuous
spectrum: $H\left\vert \omega \right\rangle =\omega \left\vert \omega
\right\rangle $, $\omega \in \left[ 0,\infty \right) $. Thus, the three
steps are:\bigskip

\textbf{First step: Selection of observables}. At $t=0$, a generic
observable can be written as%
\begin{equation}
\hat{O}(0)=\int_{0}^{\infty }\int_{0}^{\infty }\widetilde{O}(\omega ,\omega
^{\prime })|\omega ,\omega ^{\prime })d\omega d\omega ^{\prime }
\label{DA-2-1}
\end{equation}%
where $\widetilde{O}(\omega ,\omega ^{\prime })$ is any core or
distribution. We will consider only the \textit{van Hove observables} \cite%
{vanHove1957} \cite{vanHove1959}, which have a core $\widetilde{O}(\omega
,\omega ^{\prime })$ of the form:
\begin{equation}
\widetilde{O}_{vH}(\omega ,\omega ^{\prime })=O(\omega )\delta (\omega
-\omega ^{\prime })+O(\omega ,\omega ^{\prime })  \label{DA-2-2}
\end{equation}%
where $O(\omega ,\omega ^{\prime })$ is a regular function. Therefore, the
van Hove observables have the form:
\begin{equation}
\hat{O}_{vH}(0)=\int_{0}^{\infty }O(\omega )|\omega )\,d\omega
+\int_{0}^{\infty }\int_{0}^{\infty }O(\omega ,\omega ^{\prime })|\omega
,\omega ^{\prime })\,d\omega d\omega ^{\prime }  \label{DA-2-3}
\end{equation}%
These observables belong to \textit{van Hove space} $\mathcal{O}_{VH}$,
whose basis is $\left\{ \left\vert \omega \right) ,\left\vert \omega ,\omega
^{\prime }\right) \right\} $. This restriction on the observables does not
diminish the generality of SID, because the observables not belonging to the
van Hove space are not accessible to experiments \cite{SIDEXP}. The states $%
\hat{\rho}$, which do not evolve in the Heisenberg picture, are represented
by linear functionals on $\mathcal{O}_{VH}$, that is, they belong to the
dual space $\mathcal{O}_{VH}^{\prime }$ and can be written as:
\begin{equation}
\hat{\rho}=\int_{0}^{\infty }\rho (\omega )(\omega |\,d\omega
+\int_{0}^{\infty }\int_{0}^{\infty }\rho (\omega ,\omega ^{\prime })(\omega
,\omega ^{\prime }|\,d\omega d\omega ^{\prime }  \label{DA-2-4}
\end{equation}%
where $\{\left( \omega \right\vert ,\left( \omega ,\omega ^{\prime
}\right\vert \}$ is the co-basis of $\left\{ \left\vert \omega \right)
,\left\vert \omega ,\omega ^{\prime }\right) \right\} $, that is, the basis
of $\mathcal{O}_{VH}^{\prime }$. States must satisfy the usual requirements,
i.e., $\rho (\omega )$ is real and positive, and $\int_{0}^{\infty }\rho
(\omega )d\omega =1$. It is also required that $\rho (\omega ,\omega
^{\prime })$ be a regular function. Under these conditions, the states
belong to a convex set $S\subset \mathcal{O}_{VH}^{\prime }$.

According to the Heisenberg picture, the evolution of the observables is
given by%
\[
\hat{O}(t)=e^{i\hat{H}t}\hat{O}(0)e^{-i\hat{H}t}
\]%
Then, expression (\ref{DA-2-3}) becomes
\begin{equation}
\hat{O}_{vH}(t)=\int_{0}^{\infty }O(\omega )|\omega )\,d\omega
+\int_{0}^{\infty }\int_{0}^{\infty }O(\omega ,\omega ^{\prime })e^{i(\omega
-\omega ^{\prime })t}|\omega ,\omega ^{\prime })\,d\omega d\omega ^{\prime }
\end{equation}

Now we select a subset of the van Hove space, defined by the commutators.
The commutator of any two observables $\hat{O}_{1}(t)\in \mathcal{O}_{VH}$
and $\hat{O}_{2}(t)\in \mathcal{O}_{VH}$ is
\[
\left[ \hat{O}_{1}(t),\hat{O}_{2}(t)\right] =\hat{O}_{1}(t)\hat{O}_{2}(t)-%
\hat{O}_{2}(t)\hat{O}_{1}(t)
\]%
By operating with patience, we obtain:
\[
\hat{C}(t)=\left[ \hat{O}_{1}(t),\hat{O}_{2}(t)\right] =\int_{0}^{\infty
}\int_{0}^{\infty }C(\omega ,\omega ^{\prime })e^{i(\omega -\omega ^{\prime
})t}(\omega ,\omega ^{\prime }|\,d\omega d\omega ^{\prime }
\]%
where%
\[
C(\omega ,\omega ^{\prime })=\int_{0}^{\infty }\left( O_{1}(\omega ,\tilde{%
\omega}^{\prime })O_{2}(\tilde{\omega}^{\prime },\omega ^{\prime
})-O_{2}(\omega ,\tilde{\omega}^{\prime })O_{1}(\tilde{\omega}^{\prime
},\omega ^{\prime })\right) d\tilde{\omega}^{\prime }
\]%
It is important to notice that $\hat{C}(t)\notin \mathcal{O}_{VH}$\ because
it is not an Hermitian operator. However, the observable $\hat{D}(t)=i\hat{C}%
(t)\in \mathcal{O}_{VH}$\ is a legitimate quantum observable, to which we
may have empirical access.

The observable $\hat{D}(t)$ allows us to measure the degree of
incompatibility between the observables $\hat{O}_{1}(t)$ and $\hat{O}_{2}(t)$%
. For example, it can be the observable that measures the contrast between
the interference fringes in the double slit experiment. This contrast
indicates that the observable that measures by which slit the particle
passes is incompatible with the observable that measures where on the screen
the particle impacts. Then, the relevant observables considered here are the
observables $\hat{D}(t)$.\bigskip

\textbf{Second step: The computation of the expectation value}. We consider
the observable $\hat{D}$ at $t=0$
\[
\hat{D}(0)=i^{-1}\left[ \hat{O}_{1}(0),\hat{O}_{2}(0)\right]
=i^{-1}\int_{0}^{\infty }\int_{0}^{\infty }C(\omega ,\omega ^{\prime
})(\omega ,\omega ^{\prime }|\,d\omega d\omega ^{\prime }
\]%
Then we assume that the initial commutator is not $0$, i.e. $\hat{O}_{1}(0)$
and $\hat{O}_{2}(0)$\ do not commute%
\[
\hat{C}(0)=\left[ \hat{O}_{1}(0),\hat{O}_{2}(0)\right] \neq 0\longrightarrow
\hat{D}(0)\neq 0
\]%
This means that $C(\omega ,\omega ^{\prime })$ is a nonzero function. The
expectation value of $\hat{D}(t)$ is%
\[
\left\langle \hat{D}(t)\right\rangle _{\rho }=Tr\left( \rho \hat{D}%
(t)\right)
\]%
that is%
\[
\left\langle \hat{D}(t)\right\rangle _{\rho }=\left\langle i^{-1}\left[ \hat{%
O}_{1}(t),\hat{O}_{2}(t)\right] \right\rangle _{\rho
}=i^{-1}\int_{0}^{\infty }\int_{0}^{\infty }\rho (\omega ,\omega ^{\prime
})C(\omega ,\omega ^{\prime })e^{i(\omega -\omega ^{\prime })t}\,d\omega
d\omega ^{\prime }
\]%
\bigskip

\textbf{Third step: The evolution of the expectation value}. We assume that $%
\rho (\omega ,\omega ^{\prime })C(\omega ,\omega ^{\prime })$ is a regular
function, indeed simply a $L_{1}$ function in the variable $\nu =\omega
-\omega ^{\prime }$; then, the Riemann-Lebesgue theorem can be applied.
Consequently,%
\begin{equation}
\lim_{t\rightarrow \infty }\left\langle \hat{D}(t)\right\rangle _{\rho }=0
\label{DA-2-6'}
\end{equation}%
This means that, when $t\rightarrow \infty $, the expectation value of the
commutator between $\hat{O}_{1}(0)$ and $\hat{O}_{2}(0)$ becomes zero.
Therefore, the Heisenberg uncertainty relation becomes undetectable from the
experimental viewpoint.

In other words, when $t\rightarrow \infty $ we can compute the expectation
value of $\hat{D}(t)$ for any $\rho $ as follows. We may think that the
observable $\hat{D}(t)$ is a final fixed observable $\hat{D}(\ast )$ such that%

\begin{equation}
\lim_{t\rightarrow \infty }\langle \hat{D}(t)\rangle _{\rho }=\langle \hat{D}%
(\ast )\rangle _{\rho }
\end{equation}%
where $\hat{D}(\ast )=0$. This result can also be expressed as a weak limit:%
\begin{equation}
W-\lim_{t\rightarrow \infty }\hat{D}(t)=0
\end{equation}

In this way we arrive closer to the classical limit. Interference is not the
only quantum feature that vanishes:  from the experimental viewpoint, the
initially non-commuting observables, tend to commute after a sufficient time.

\section{Classical limit in the logical structure}

The central task of this work is to study the classical limit of quantum
mechanics from the point of view of the logical structure of the theory. We
have already seen that the essential difference between the lattice of the
classical properties and the lattice of the quantum properties is that in
the first one the distributive equalities hold. Only in a distributive and
ortocomplemented lattice we have a Boolean structure of the properties.

Therefore, the study of the classical limit requires the study of under what
conditions a quantum structure of properties becomes Boolean.  It is clear
that this limit must involve a non-unitary evolution, a coarse-grained \cite%
{Formal}, or some additional element; otherwise, a set of properties whose
projectors do not commute, and therefore form a non-classical algebra, will
never lose this feature.  But we have seen that
evolutions of this type are involved in the search of the classical limit as
a result of the mechanism of decoherence .

The decoherence studied in the previous sections meets our goal. In fact, on
the basis of the evolutions studied here, it is possible to show that the
commutator between certain observables $\hat{O}_{1}(t)$ and $\hat{O}_{2}(t)$
vanishes, at least in terms of their expectation values. Therefore, if we
measure the observable $\hat{D}(t)$ at the beginning of the process, its
expectation value is not zero; but if we measure this observable at the end
of the process, its expectation value is almost zero. This means that, from
the observational point of view, we can assume that $\hat{O}_{1}$ and $\hat{O%
}_{2}$ are compatible observables. But, does this mean that now we have
recovered distributivity?

We can interpret the evolution of these observables as follows. Let us
consider two properties, $A$ corresponding to the value $o_{1}$ of the
observable $\hat{O}_{1}$, and $B$ corresponding to the value $o_{2}$ of the
observable $\hat{O}_{2}$. If we think these observable properties
 as vectors in the Hilbert space, then they enclose an angle.  The evolution is such that the angle between the
vectors representing the properties \textquotedblleft $o_{1}$%
\textquotedblright\ and \textquotedblleft $o_{2}$\textquotedblright\ gets
smaller. While the angle is not exactly zero, we have non-distributivity.
But in the infinite time limit, the angle between the vectors representing
the properties $A$ and $B$ becomes zero, and the corresponding observables
turn out to commute with each other. Therefore, distributivity is recovered.

In other words, decoherence can be also viewed as a process that turns
incompatible observables into compatible observables and, as a consequence,
that turns the quantum logic into the classical Boolean logic.

\section{Conclusion}

Through the decoherence of the expectation values it is possible to study
the classical limit of a quantum system in terms of decohering observables.
These are the relevant observables of the system, and from their property
values it is possible to construct the logical structure of interest.

This endows decoherence with a semantic content stronger than that involving
the mere process by which the interference terms vanish. The evolution of
the commutators allows us to understand decoherence as a process by which
the logical structure of what can be said about the system acquires
classical characteristics, i.e., becomes Boolean. These features have
relevant consequences on the calculation of the probabilities of the values
of the decohering observables.

Therefore, we can establish the transition between two logics, quantum logic
and classical logic, from the observational point of view. We propose to
continue this line of work by studying the evolution of the logical
properties of the system in time, for example, by analyzing the evolution of
the observables, not of their the expectation values. On the other hand,
although the usual lattice is constructed from properties, we can try to
build a lattice from expectation values. In both cases, we could describe
the transition from quantum logic to Boolean logic.



\begin{thebibliography}{99}




\bibitem{Bub1997} Bub, J.: Interpreting the Quantum World, Cambridge:
Cambridge University Press (1997).


\bibitem{Zurek1981} Zurek, W.:  Pointer basis of
quantum apparatus: into what mixtures does the wave packet
collapse?, Physical Review D, 24: 1516-25 (1981).

\bibitem{Zurek2003} Zurek, W.:  Decoherence,
einselection, and the quantum origins of the classical,
Reviews of Modern Physics, 75: 715-76 (2003).


\bibitem{ZurekPaz2002}  Paz, J. P.,   Zurek, W. H.:
Environment-induced decoherence and the transition from quantum to
classical, en D. Heiss (ed.),Lecture Notes in Physics, Vol. 587, Springer, Heidelberg (2002).




\bibitem{Schlosshauer2007} Schlosshauer, M.: Decoherence and the
Quantum-to-Classical Transition, Berlin: Springer   (2007).

\bibitem{GTFD} Castagnino, M.,  Fortin, S., Laura, R.,  Lombardi, O.:
Classical And Quantum Gravity, \textbf{25}, 154002 (2008).


\bibitem{EPSA2009} Lombardi O., Fortin, S.,  Castagnino M.:
 The problem of identifying the system and the environment
in the phenomenon of decoherence, in H. W. de Regt, S.
Hartmann and S. Okasha (eds.), European Philosophy of Science Association
(EPSA). Philosophical Issues in the Sciences Volume 3, Berlin: Springer, pp.
161-174 (2012).



\bibitem{Formal} Castagnino M., Fortin S.:  International
Journal of Theoretical Physics, \textbf{52} (5) , pp. 1379-1398 (2013).

\bibitem{SID7} Castagnino, M., Lombardi, O.: Int. Jour.
Theor. Phys., \textbf{42}, 1281 (2003).

\bibitem{fortinlombardi2014}Fortin S., Lombardi O.: Foundations of Physics, in press, DOI:10.1007/s10701-014-9791-3 (2014).


\bibitem{BirkhoffNeumann1936}  Birkhoff,  G.,  von Neumann, J.: The logic of
quantum mechanics, Annals of Math. \textbf{37}, pp. 823-843 (1936).




\bibitem{CohenLibro}  Cohen, D. W.: An introduction to Hilbert space
and quantum logic, Springer-Verlag, New York (1989).

\bibitem{KieferPolarski2009}  Kiefer C.,  Polarski D.: Adv. Sci. Lett., \textbf{2}, 164-173 (2009).

\bibitem{MittelstaedtLibroLogica}  Mittelstaedt, P.: Quantum Logic.
D. Reidel Publishing Company, Dordrecht (1978).



\bibitem{HughesLibro} Hughes,  R.I.G.: The Structure and Interpretation
of Quantum Mechanics, Harvard Univerity Press, Cambridge (1992).

\bibitem{Holik} F. Holik, F.,  Plastino, A., Saenz, M.: Annals Of Physics, Volume 340, Issue 1, 293-310, (2014).


\bibitem{Redei}  Redei, M.,   Summers, S.:  Stud. Hist. Philos. Sci. B Stud. Hist. Philos. Modern Phys.  \textbf{38} (2),  390-417, (2007).


\bibitem{Gudder}  S.P. Gudder, S.P.:  Stochastic Methods in Quantum Mechanics, North Holland, New York-Oxford, (1979).


\bibitem{OmmesLibroInterpretacion}  Omn\`{e}s, R.: The interpretation
of quantum mechanics, Princeton University Press (1994).

\bibitem{Boole1854} Boole, G.: An Investigation of the Laws of
Thought, on Which are Founded the Mathematical Theories of Logic and
Probabilities, London: Macmillan  (1854).



\bibitem{Sakurai}  Sakurai, J. J.: Modern Quantum Mechanics, Revised
Edition, Addison-Wesley, New York (1994).



\bibitem{Zurek1982}  Zurek, W. H.: Phys. Rev. D, \textbf{26}, 1862
(1982).

\bibitem{KieferPolarski1998}  Kiefer, C.,  PolarskiD.: Annalen Phys. 7,
137-158 (1998).


\bibitem{SID1} Castagnino, M.,  Laura, R.: Phys. Rev. A,
\textbf{56,} 108 (1997).

\bibitem{SID2}  Laura, R., Castagnino, ;.: Phys. Rev.
\textit{A}, \textbf{57}, 4140 (1998).

\bibitem{SID3} Laura, R.,  M. Castagnino, M.: Phys. Rev. E, \textbf{57},
3948 (1998).

\bibitem{SID4}  Castagnino, M.: Int. Jour. Theor. Phys.,\textbf{38}, 1333 (1999).

\bibitem{SID5}  Castagnino, M., Laura, R.:  Phys. Rev. A, \textbf{62},
022107 (2000).

\bibitem{SID6} Castagnino, M.,  Laura, R.: Int. Jour. Theor. Phys.,
\textbf{39}, 1767  (2000).



\bibitem{SID8} Castagnino, M.,  Ordo\~{n}ez, A.:  Int. Jour. Theor.
Phys., \textbf{43}, 695 (2004).

\bibitem{SID9} Castagnino, M.:Physica A, \textbf{335},
511 (2004).

\bibitem{SID10} Castagnino, M.: Phys. Lett. A, \textbf{357}, 97,
(2006).

\bibitem{SID11}  Castagnino, M.:  Physica A, \textbf{335},
511 (2004).

\bibitem{SID12}  Castagnino, M., Lombardi, O.:  Phil. Scie.,\textbf{72}
,764 (2005).

\bibitem{SID13}  Castagnino, M.,  Gadella, M.: Found. Phys., \textbf{36}, 920 (2006).


\bibitem{SIDDiscreto} Castagnino M., Fortin S.:  International
Journal of Theoretical Physics, \textbf{50 }(7) , pp. 2259-226 (2011)





\bibitem{Brasilero} Fortin S., Lombardi O.,  Castagnino M.: %
Brazilian Journal of Physics, \textbf{44}, pp. 138-153, (2014).

\bibitem{Antoniou}  Antoniou, I.,  Laura, R.,  Tasaki S., and  Suchaecki, z.: Physica A, 241, 737-772 (1997).


\bibitem{vanHove1957} van Hove, L.: Physica \textbf{23}, 441 (1957)

\bibitem{vanHove1959} van Hove, L.: Physica \textbf{25}, 268 (1959)



\bibitem{SIDEXP}  Castagnino, M.,  Lombardi, O.:  Studies In History
and Philosophy of Modern Physics, \textbf{35}, pp. 73-107 (2004).













\end{thebibliography}


\end{document}